\documentclass[twocolumn]{aastex62}

\usepackage{apjfonts}
\usepackage[T1]{fontenc}
\usepackage{color}
\usepackage{amsmath,amstext}
\usepackage{hyperref}
\usepackage{natbib}

\newcommand\beq{\begin{equation}}
\newcommand\eeq{\end{equation}}

\newcommand{\dd}{\mathrm{d}}
\newcommand{\diff}[2]{\frac{\dd #1}{\dd #2}}

\newcommand{\alphaOneSigma}{2.35^{+0.55}_{-0.36}}

\submitjournal{ApJL}

\shorttitle{}
\shortauthors{Perna et al.}

\begin{document}
\title{CONSTRAINING THE BLACK HOLE INITIAL MASS FUNCTION WITH LIGO/VIRGO OBSERVATIONS}

\author{Rosalba Perna}
\affiliation{Department of Physics and Astronomy, Stony Brook University, Stony Brook, NY, 11794, USA}
\affiliation{Center for Computational Astrophysics, Flatiron Institute, 162 5th Avenue, New York, NY 10010, \
USA}

\author{Yi-Han Wang}
\affiliation{Department of Physics and Astronomy, Stony Brook University, Stony Brook, NY, 11794, USA}

\author{Will M. Farr}
\affiliation{Department of Physics and Astronomy, Stony Brook University, Stony Brook, NY, 11794, USA}
\affiliation{Center for Computational Astrophysics, Flatiron Institute, 162 5th Avenue, New York, NY 10010, \
USA}

\author{Nathan Leigh}
\affiliation{Department of Physics and Astronomy, Stony Brook University, Stony Brook, NY, 11794, USA}
\affiliation{Departamento de Astronom\'ia, Facultad de Ciencias F\'isicas y Matem\'aticas,
Universidad de Concepci\'on, Concepci\'on, Chile}
\affiliation{Department of Astrophysics, American Museum of Natural History, Central Park West and 79th Street, New York, NY 10024}

\author{Matteo Cantiello}
\affiliation{Center for Computational Astrophysics, Flatiron Institute, 162 5th Avenue, New York, NY 10010, \
USA}
\affiliation{Department of Astrophysical Sciences, Princeton University, Princeton, NJ 08544, USA}

\correspondingauthor{Rosalba Perna}
\email{rosalba.perna@stonybrook.edu}

\begin{abstract}

Prior to the detection of black holes (BHs) via the gravitational waves (GWs)
they generate at merger, the presence of BHs was inferred in X-ray binaries,
mostly via dynamical measurements, with masses in the range between $\sim
5-20~M_\odot$. The LIGO discovery of the first BHs via GWs was surprising in
that the two BHs that merged had masses of $35.6^{+4.8}_{-3.0}$ and
$30.6^{+3.0}_{-4.4}\,M_\odot$, which are both above the range inferred from
X-ray binaries. {  With 20 BH detections from the O1/O2 runs,  the
distribution of masses remains generally higher than the X-ray inferred one,
while the effective spins	are generally lower, suggesting that, at least	in
part, the GW-detected population might be of dynamical origin rather than
produced by the common evolution of field binaries. } Here we perform
high-resolution N-body simulations of a cluster of isolated BHs with a range of
initial mass spectra {  and upper mass cut-offs, and study the resulting
binary mass spectrum resulting from the dynamical interactions.  Our clusters
have properties similar to those of the massive remnants in an OB association
$\sim 10 \, \mathrm{Myr}$ after formation.  We perform a likelihood analysis for
each of our dynamically-formed binary population against the data from the O1
and O2 LIGO/Virgo runs. We find that an initial mass spectrum $M_{\rm BH}\propto
M^{-2.35}$ with an upper mass cutoff $M_{\rm max}\sim 50M_\odot$ is favored by
the data, together with a slight preference for a merger rate that increases
with redshift. }

\end{abstract}

\keywords{gravitational waves ---  black hole physics --- methods: numerical --- binaries: general}

\section{Introduction}

The existence of black holes (BHs) is one of the primary predictions of the Theory of General Relativity.
Prior to their direct discovery via the gravitational waves they generated in a merger event
\citep{Abbott2016}, their presence was inferred via dynamical mass measurements in X-ray binaries (see i.e. \citealt{Witz2014} for a summary).
The values of the inferred masses vary between $\sim 4-5 M_\odot$ to about 20~$M_\odot$,
marking a clear separation with the inferred neutron star masses, for which the largest
measurement to date is 1.96 $M_\odot$ \citep{Demorest2010}.

The discovery of the first binary black hole merger via the gravitational waves
generated at the time of coalescence led to a mass measurement for the BH
components of the merging binary: $35.6^{+4.8}_{-3.0}$ and
$30.6^{+3.0}_{-4.4}\,M_\odot$.  The large BH masses, both well above the maximum
value measured to date in X-ray binaries, came as a surprise
\citep{GW150914Astrophysics}.  The discovery triggered an intense debate in the
literature on the formation pathway of this BH binary.

Broadly speaking, most formation avenues can be classified within one
of two channels: isolated binary evolution, in which two massive stars
evolve till their death while remaining gravitationally bound \citep[e.g. ][]{Podsiadlowski2003,Dominik2013, Belczynski2016,Merchant2016,Mandel2016}, and dynamical
formation by gravitational capture in dense environments, where
binaries are being formed from isolated BHs as a result of frequent
dynamical interactions \citep[e.g.][]{Sigurdsson1993,Portegies2000,OLeary2006,Miller2009,Mapelli2013,Leigh2014,Ziosi2014,Rodriguez2016,Antonini2016,Chatterjee2017, Rodriguez2018,Samsing2018,Samsing2018a,Generozov2018,Antonini2018,Banerjee2018,Fragione2018a,Fragione2018b,
Dicarlo2019,Ye2019}.
 The theoretically-predicted rates are rather uncertain for both scenarios:
the models explored by \citet{Belczynski2016} yield rates which vary between $\sim
6-1000$~Gpc$^{-3}$~yr$^{-1}$.  More recent, state-of-the art estimates
of the rates of dynamical formation in Globular Clusters yield a range
of $4-18$~Gpc$^{-3}$~yr$^{-1}$\citep{Rodriguez2018}. Both these rates are compatible with
the current observationally-determined value by LIGO, of
9.7-101~Gpc$^{-3}$~yr$^{-1}$ \citep{Abbott2019}.
Both channels can in principle contribute to the observed population, something that can be tested since many more mergers are going to be detected in the future.

To date, after the first two observing runs of LIGO/Virgo, there have been 10
BBH mergers reported \citep{Abbott2019}. While the smallest masses (4 out of 20)
fall within the upper range of the masses inferred for X-ray binaries,
the other 16 are all bigger, with the largest being $50.6^{+16.6}_{-10.2}
M_\sun$. While the distributions are clearly not disjoint,
there is a marked preference in GW-detected BHs for larger
masses than in those found via X-ray binaries, raising the question of whether
the two observed populations are dominated by the same progenitor population.

An independent piece of evidence which raises the same question  is constituted by the measured
spins\footnote{ What GWs measure is the so-called effective spin, i.e. the mass-weighed projection of the spins onto the orbital angular momentum. }  (see i.e. \citealt{Farr2017, Roulet2019}): generally
high and aligned with the orbital angular momentum in the XRBs (especially the persistent ones), and typically low and
isotropic in the GW-detected BHs. This trend is consistent with studies  suggesting that,
while isolated binaries preferentially yield BHs with spins aligned with the orbital angular momentum (i.e. \citealt{Kalogera2000}),
dynamically formed BHs have no preferred direction for alignment (i.e. \citealt{Portegies2000}).

{In this paper we investigate the possibility that the observed BH
population  is dominated by a dynamical formation channel complementary to the dynamical channel discussed in the context of globular clusters. More specifically,  we generate small
clusters of BHs with properties appropriate for the massive remnants of an OB
association $\sim 10 \, \mathrm{Myr}$ after its formation, following the
supernovae explosions of its massive constituents. { Therefore, albeit dynamically formed, the binary population discussed here does not necessarily need to originate in globular clusters, but could also be associated with field stars.}
Using high-resolution N-body
simulations of these clusters with a range of mass spectra, we explore the
dependence of the BH binary mass distribution on the mass spectrum and upper
mass cut-off of the individual BHs (\S2). We perform a statistical comparison
with the data from the O1 and O2 LIGO/Virgo runs to study the consistency
between the data and the simulation results, and assess statistical preferences
towards an initial BH mass spectrum (\S3). We summarize and conclude in \S4.}

\section{The BBH mass spectrum from dynamical interactions}

Motivated by the considerations of Sec.~1, here we set to
perform high-resolution N-body simulations of a
cluster of BHs, with the goal of exploring the dependence of the mass
spectrum and the orbital parameters of the dynamically-formed binaries
on the mass spectrum of the isolated BH population.
We note that the mass spectrum of BH binaries produced as a result of dynamical
interactions in clusters has a long history in the literature,
predating the era of gravitational waves. A preferential tendency for
dynamically formed binaries to have heavier mass has been noted in a
number of works \citep{OLeary2006,Miller2009,Ryu2016,Rodriguez2018a,Dicarlo2019}.
{ Our fewbody simulations, while not including the effects of the cluster potential, they
allow us to accurately follow binary formation, since in dense environments this is dominated by 3-body scatterings,
as well as perform a large number of Monte Carlo realizations. }

\begin{figure*}
\includegraphics[width=18cm]{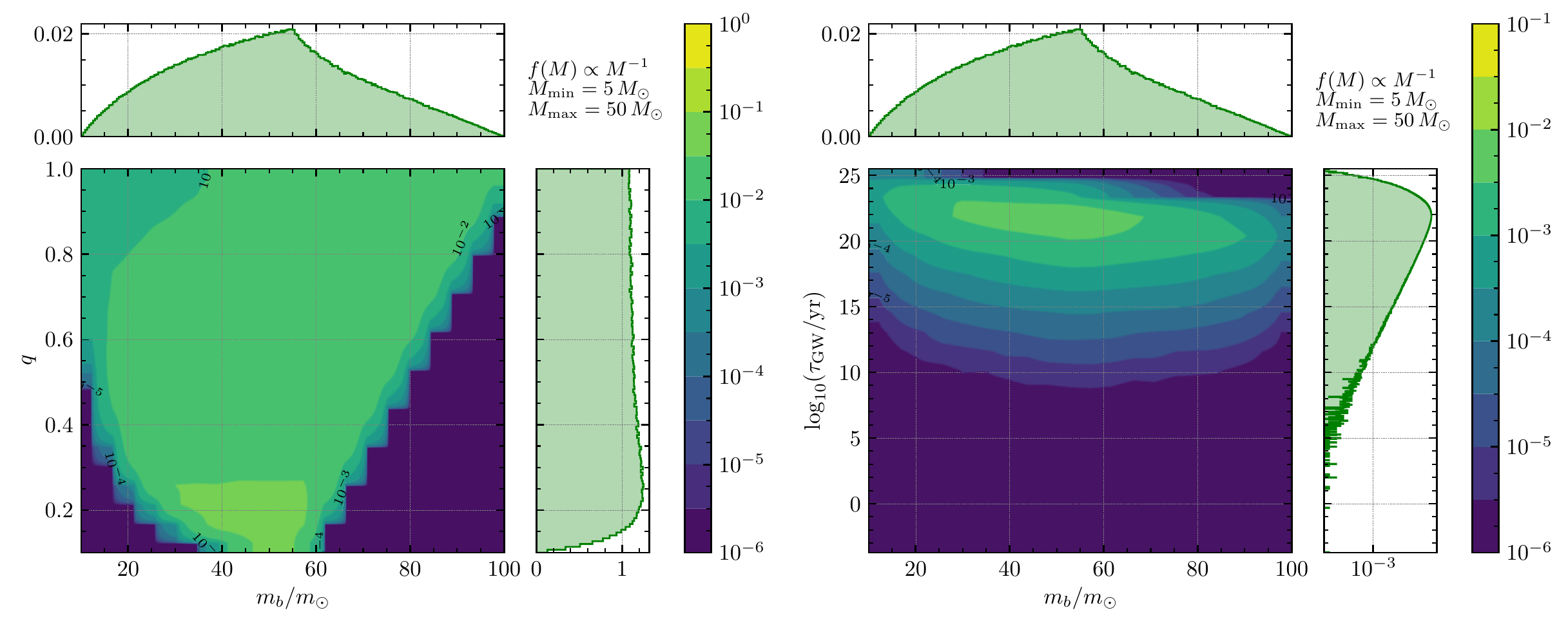}
\caption{{\em Left:} The distribution of mass ratios and total binary masses
  of dynamically-formed binaries from a cluster of BHs with a mass spectrum $\propto M^{-1}$
  between 5-50~$M_\odot$. {\em Right:} The corresponding merger times as a
  function of the total binary masses. In both panels, the top and the right plots display the collapsed 2D distributions onto the corresponding axis.}
\label{fig:BHflat}
\end{figure*}

\begin{figure*}
\includegraphics[width=18cm]{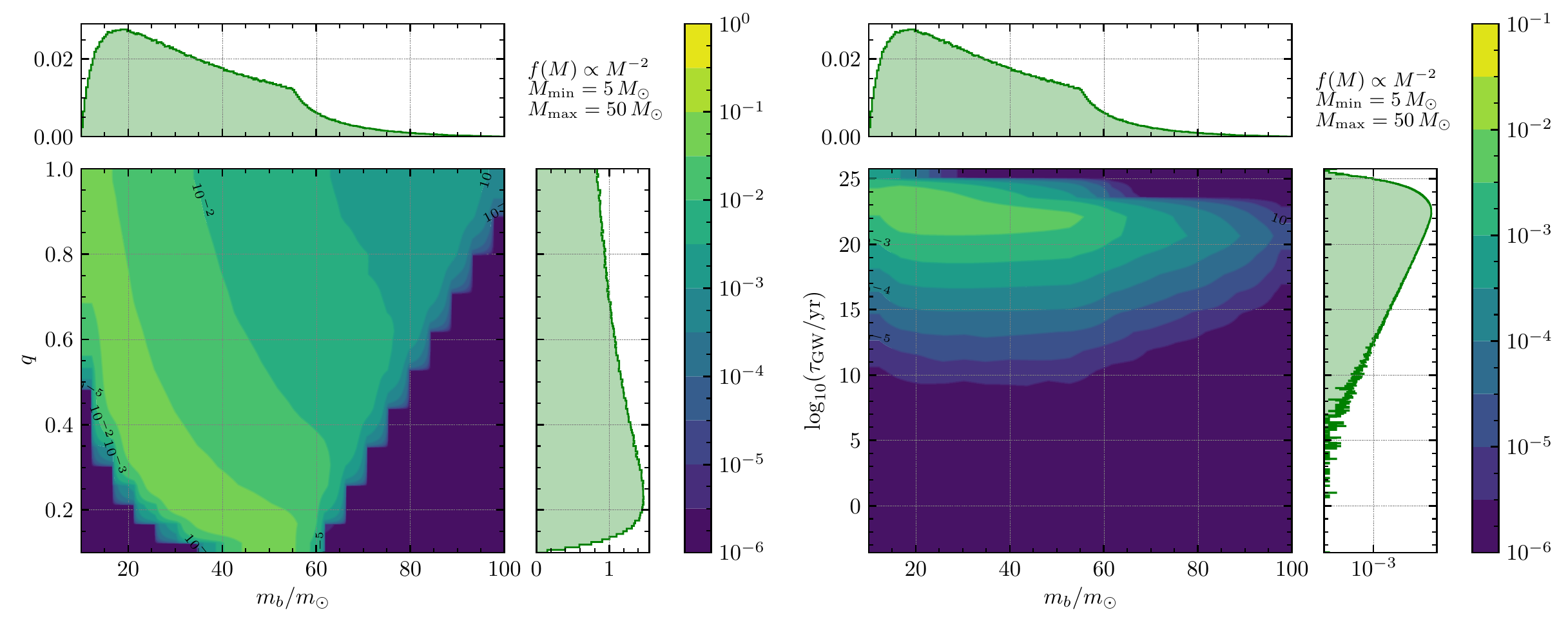}
\caption{Same as in Fig.\ref{fig:BHflat} but for a power-law mass spectrum of index -2
for the interacting isolated BHs.}
\label{fig:BH1}
\end{figure*}

We consider a cluster of 20 BHs, with an initial binary fraction equal
to zero in order to purely explore the binary properties due to
dynamical formation. Their positions in the cluster are initially distributed randomly
in a sphere of radius 0.1~pc. { Astrophysically, this configuration can be thought of as representing
the remnants of an OB association (typically comprised of $\sim 10-100$ stars), still confined within  { the dense core of a molecular cloud\footnote{ Dynamically, this population however bears resemblance with a core-collapsed cluster (see e.g. \citealt{Sigurdsson1993,Rodriguez2016}).} (e.g. \citealt{Zhou1994})}. From
a numerical point of view, we note that the particular number of 20 was chosen
as a 'sweet spot': large enough for obtaining a reasonably well sampled binary mass distribution,
but small enough to enable the running of a large number of realizations with high numerical accuracy. However,
we ran several additional simulations with
different numbers of BHs in order to  verify that the shape of the
binary distribution remains statistically the same as the number of BHs is varied. Similarly,
we chose the size of the initial spatial domain after verifying that it was large enough that the results
for the binary mass distributions were converged as the region size was varied.}

\begin{figure*}
\includegraphics[width=18cm]{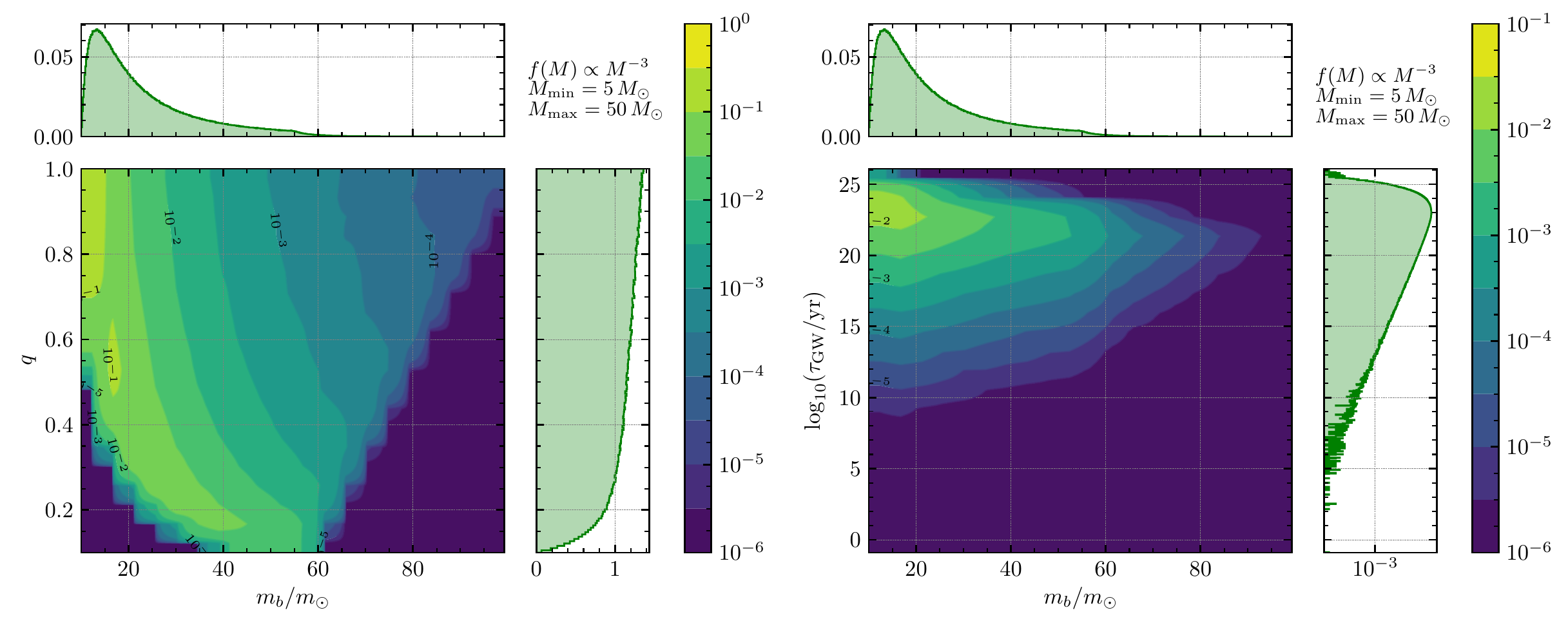}
\caption{Same as in Fig.\ref{fig:BHflat} but for a power-law mass spectrum of index -3
for the interacting isolated BHs.}
\label{fig:BH2}
\end{figure*}

The BHs were assigned a mean speed of 5~km/s, as typical of the
velocity dispersions observed in low-mass star clusters
\citep[e.g.][]{Harris1996}. We follow their evolution using our code
\href{https://github.com/YihanWangAstro/Template-SpaceX}{\tt SpaceHub} (see \citealt{Wang2019} for details), which employs
the \textit{ARCHAIN} algorithm \citep{Mikkola2008} to accurately trace
the motion of tight binaries with arbitrarily large mass ratios and
eccentricities, and a chain structure to reduce the round-off errors
from close encounters. {
  Binaries are detected in the simulations when the following
  conditions are verified: (a) the BH masses $m_1$ and $m_2$ are
  gravitationally bound to one another; (b) the system $(m_1+m_2)$ is
  gravitationally unbound in the potential of the remaining BHs; (c)
  the binary has traveled a large enough distance from the original BH
  cluster. We chose the distance to be 20 times the size of the
  original cluster, after verifying that the simulation results are
  converged for that value.}  The initial BH masses are drawn from a
distribution with $M_{\rm min}=5 M_\odot$ and $M_{\rm max}$ {  varying within the range
$40-50\,M_\odot$ in steps of $5\,M_{\rm \odot}$}.    The mass spectrum of BH
  remnants, in addition to depending on the
  evolutionary model, is also strongly dependent on metallicity,
  varying from an almost flat distribution at solar metallicities to
  an almost linear dependence on the main sequence mass at low
  metallicities and for high-mass progenitor stars (e.g.  \citealt{Spera2015}).  {  Therefore, we explored a variety
  of mass spectra, ranging from a flat distribution to a power-law
  $M^{-\alpha}$  with index $\alpha=4$ in steps of 0.5. Additionally,  we also investigated the particular case of $M^{-2.35}$, corresponding to a BH mass spectrum reflective of the initial mass function of the massive  progenitor stars \citep{Salpeter1955}. }

  We show some representative
  results in Fig.~\ref{fig:BHflat} ($\alpha=1$), Fig.~\ref{fig:BH1}
($\alpha=2$), and Fig.~\ref{fig:BH2} ($\alpha=3$), all with $M_{\rm max}=50\,M_\odot$.
In the left panels
we show the mass ratio $q$, while in the right panels the merger time
$\tau_{\rm GW}$ calculated according to Peters' formulae
\citep{Peters1964}, both versus the total mass of the binaries.
The
flatter mass spectrum $(\alpha=1)$ yields a large fraction of of binaries with total
mass $\sim 40-65\,M_\odot$.
Interestingly, not only is there an
exceedingly large number of very massive binaries formed, but they are
also the ones which are more tightly bound, hence resulting in shorter
merger times due to GW emission (note a similar result found by
\citet{Ryu2016} in the context of the formation of the first X-ray
binaries).  The most massive binaries also tend to have higher mass
ratios. Note that the kink
at $\sim 55\,M_\odot$
corresponds to the mass of the binary formed by the most and the least
massive BHs, which results in
the mass distribution to undergoe a change of
slope as it passes through that point.

  As the mass spectrum steepens to $M^{-2}$, the
  distribution becomes more apparently dominated by lower-mass
  binaries, but a tail in the high mass range still remains.
  {  For $\alpha=3$, the binary mass distribution becomes clearly peaked towards low masses, and the high-mass tail of the distribution becomes vanishingly small.} { The fraction of BHs which end up as ejected binaries is also dependent on the mass spectrum. In particular, we found this fraction to be 0.11\% for $\alpha=1$, 0.09\% for $\alpha=2$ and
  0.065\% for $\alpha=3$. Correspondingly, the relative fraction of those binaries which merge within the Hubble time is 0.092\%, 0.069\%, and 0.048\% for $\alpha=1,2,3$, respectively.  }

Of particular relevance for the LIGO/Virgo results is the fact that
the more massive binaries tend to be the more tightly bound; the
heaviest objects are in fact the ones with the largest cross-sections
for encounters, hence they undergo the most scatterings and end up as
the hardest binaries.
{ This tendency is especially pronounced for
  shallower BH mass spectra, when the number of massive BHs is not much smaller than the number of lighter BHs. For very steep slopes of the mass spectrum, the interaction probability becomes dominated by the number of small BHs, which hence have much higher chances of interacting, and thus forming tight binaries.

{ 
\section{Statistical comparison with the LIGO/Virgo data from the O1/O2 runs}
\label{sec:compare-LIGO}

We now wish to compare the  models described in the previous section to the 10
binary black hole mergers observed by LIGO and Virgo in GWTC-1
\citep{Abbott2019}\footnote{Our analysis can be found at
\url{https://github.com/farr/ClusterBHGW}.}.  The analysis is very similar to
\citet{Farr2017}: here we have a collection of zero-parameter models that
predict the mass distribution of merging black holes.  Unlike in
\citet{Farr2017}, the detectability of mergers is a strong function of mass, so
we must account for selection effects; see \citet{Mandel2018} and references
therein.

We are comparing to a dataset, $\mathbf{d}$, consisting of a catalog of detections, $i = 1,
\ldots, N_\mathrm{det} = 10$.  The Bayseian posterior
probability of a particular mass model, $M$, is given by
\begin{equation}
p\left( M \mid \mathbf{d} \right) \propto p\left( \mathbf{d} \mid M \right) p\left( M \right).
\end{equation}
$p\left( \mathbf{d} \mid M \right)$ is sometimes known as the "Bayes
factor;" it is the likelihood of model $M$ given the observed data.
$p(M)$ is the prior probability of model $M$, which we are free to
assign based on our experience and intuition; we describe our model priors below as we discuss Figure \ref{fig:allmergers}.

We assume that the noise realization in the LIGO and Virgo detectors is statistically independent for each event, so that
\begin{equation}
p\left( \mathbf{d} \mid M \right) = \prod_{i=1}^{N_\mathrm{det}} p\left( d_i \mid M \right).
\end{equation}
Each model makes predictions about the masses of the merging binaries.  In principle each model also makes predictions about the redshift distribution of merging binaries, but we leave study of this prediction to future work.  Instead, we select only the mergers whose time to merger is $t_\mathrm{GW} < 10^{10} \, \mathrm{yr}$ and, following \citet{Fishback2018}, impose a parameterized redshift distribution of events corresponding to a volumetric merger rate in the comoving frame of
\begin{equation}
    \diff{N}{V \dd t} \propto \left( 1 + z \right)^{\lambda}.
\end{equation}
Setting $\lambda = 3$ gives a merger rate that approximately tracks the star formation rate; setting $\lambda = 0$ gives a merger rate that is constant in the comoving frame \citep{Fishback2018}.
The likelihood of the data depends on the masses and redshifts of the merging systems, which are subject to selection effects, so we have \citep{Farr2017,Mandel2018}
\begin{multline}
p\left( d_i \mid M \right) = \\ \frac{\int \, \dd m_1 \, \dd m_2 \, \dd z \, p\left( d_i \mid m_1, m_2, z \right) p\left( m_1, m_2, z \mid M \right)}{\int \, \dd m_1 \, \dd m_2 \, \dd z \, P_\mathrm{det}\left( m_1, m_2, z \right) p\left( m_1, m_2, z \mid M \right)}.
\end{multline}
Here the numerator is the likelihood of the LIGO data given the masses
and redshifts predicted by the model $M$, and the denominator is the correction for
the selection function and gives the average detectability for the model population.

\begin{figure}
\includegraphics[width=8.5cm]{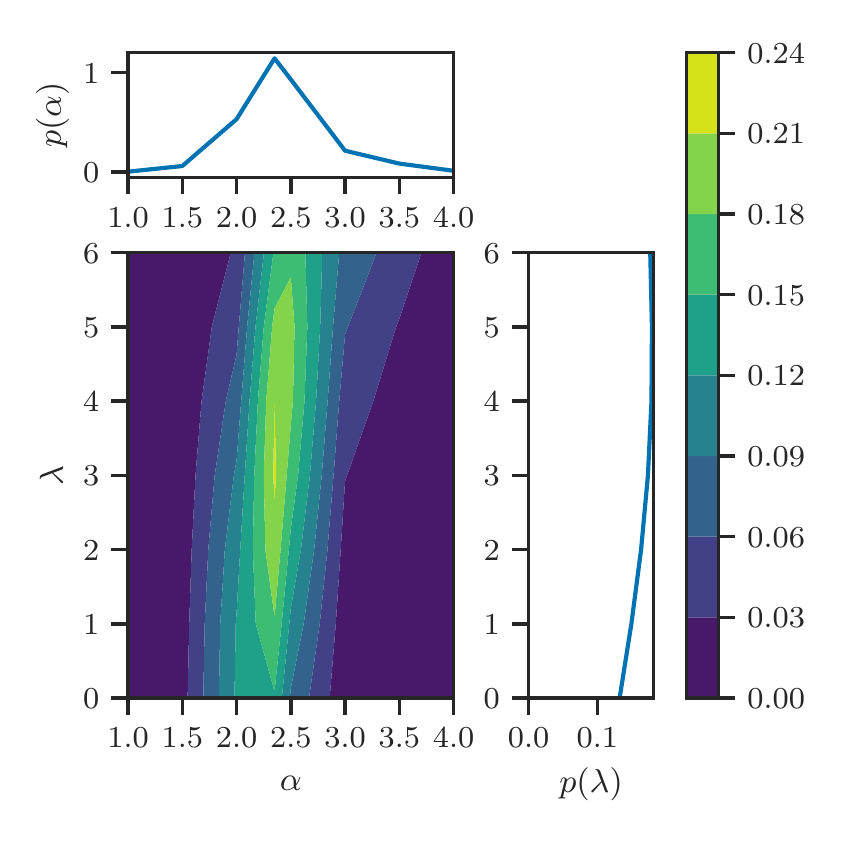}
\caption{  Two dimensional posterior on $\alpha$ (slope of the black hole mass
function) and $\lambda$ (slope of the merger rate versus redshift) inferred from
the 10 LIGO/Virgo BBH detections discussed in Section \ref{sec:compare-LIGO} and
the one-dimensional marginal posteriors for $\alpha$ and $\lambda$.  We impose a
flat prior density in $\alpha$ and $\lambda$.  The observations weakly favor a
merger rate that increases rapidly with redshift ($\lambda \simeq 3$ corresponds
to a merger rate that tracks the star formation rate
\citep{Fishback2018,Madau2014}).  The posterior is maximized (in both 1- and
2-D) at $\alpha = 2.35$, with a 1$\sigma$ (68\% credible) interval of $\alpha =
\alphaOneSigma{}$.}
\label{fig:allmergers}
\end{figure}

The denominator is independent of the data, $d_i$, and common to all events.  We
use a Monte-Carlo estimate of the integral \citep{Farr2019} obtained by
generating synthetic merger events and detecting them using an analytic estimate
of the LIGO/Virgo O1+O2 sensitivity \citep{Abbott2016c}.

The numerator can also be computed via Monte-Carlo using parameter
estimation samples from \citet{Abbott2019}.
Those are drawn from a posterior density that incorporates the
likelihood and a prior, $p_\mathrm{PE}\left( m_1, m_2, z \right)$ (where 'PE' stands for parameter estimation)
\begin{equation}
m_1, m_2, z \sim p\left( d_i \mid m_1, m_2, z \right) p_\mathrm{PE} \left( m_1, m_2, z \right),
\end{equation}
so the likelihood integral that we need can be computed via
\begin{multline}
\int \, \dd m_1 \, \dd m_2 \, \dd z \, p\left( d_i \mid m_1, m_2, z \right) p\left( m_1, m_2, z \mid M \right)  \\ \propto \left\langle \frac{p\left( m_1, m_2, z \mid M \right)}{p_\mathrm{PE}\left( m_1, m_2, z \right)} \right\rangle,
\end{multline}
where the final average is taken over the PE samples.  We use a Gaussian kernel
density estimator in a two-dimensional, un-constrained parameter space $\left(
x, y \right) = \left( \log m_1, \log m_2 - \log \left( m_1 - m_2 \right)
\right)$ to smooth the distribution over masses predicted by each model when
computing the model likelihood.

We begin our analysis by considering,  for each
of the models described in Sec.~3, a population drawn from a redshift distribution varying
from constant ($\lambda=0$), to a rapidly evolving one ($\lambda=6$).  We impose
uniform prior density in models in $\alpha$ and $\lambda$.   The 2D posterior on
models in $\alpha$-$\lambda$ space is displayed in Fig.~\ref{fig:allmergers},
together with the 1D projections on the $\lambda$ and $\alpha$ axis, for the case with $M_{\rm max}=50\,M_\odot$, which appears to provide the best match.

The analysis shows that there is a slight preference for an evolving redshift
distribution, though the trend is only marginal.  This is consistent with the
results on redshift evolution from \citet{Abbott2018}.  Marginalizing over
$\lambda$, the preferred spectral index is found to be $\alpha=2.35$.  A rough
1$\sigma$ (68\% credible) interval for $\alpha$ is $\alpha = \alphaOneSigma{}$.

In Figure \ref{fig:mass-predictions} we compare our models' predictions for the
distribution of the observed total mass of 10 merging binaries for various
values of $M_{\rm max}$, fixing $\lambda = 3$ (i.e.\ a merger rate that tracks
the star formation rate), and  $\alpha=2.35$. The predictions incorporate both
the GW selection function and also observational uncertainties. No model fully
reproduces the distribution of observed total masses, though the observations
lie within the 1$\sigma$ band of the $M_{\rm max} = 50 M_\odot$, $\alpha = 2.35$
model. We remark that our models are minimally parameterized, being dependent
only on the two parameters $\alpha$ and $M_{\rm max}$; the shape of the binary
mass distribution is then determined by dynamics alone, and hence its
resemblance to the observed distribution for some astrophysically interesting
sets of parameters ($\alpha=2.35$ and $M_{\rm max}=50\,M_\odot$) is especially
intriguing. }

\begin{figure}
\includegraphics[width=8.5cm]{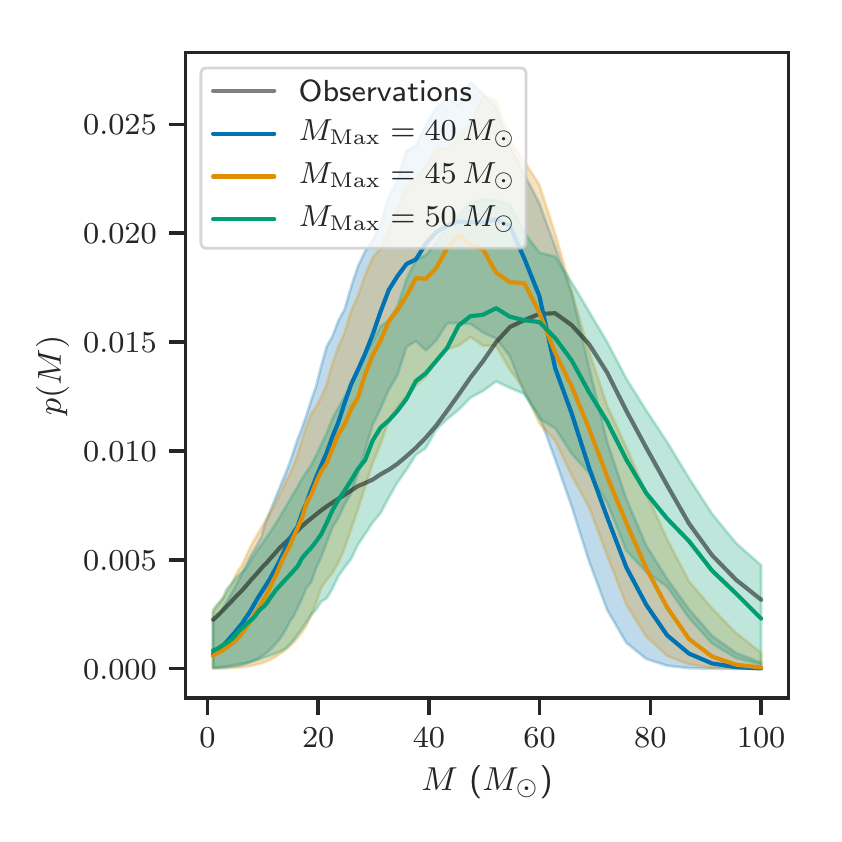}
%
\caption{\label{fig:mass-predictions} \textbf{Comparison between the model
prediction for the distribution of observed total mass after ten observations
for models at the indicated values of $M_{\rm max}$ (with $\alpha =2.35$) and
the LIGO/Virgo observations (black line).  The predictions from the models
incorporate our estimate of the LIGO/Virgo selection function
\citep{Abbott2016c} and also estimate the observational uncertainties for each
synthetic detection by matching to the LIGO/Virgo detection with the nearest
median total mass.  The solid line shows the median over 100 realizations of 10
simulated BBH detections from each model and the bands show the 1$\sigma$ (68\%
credible) interval in the total mass distribution.  None of the models fully
reproduces the precise shape of the observed total mass distribution, but the
observations remain within the 1$\sigma$ uncertainty band for the model with $M_{\rm
max} = 50M_\odot$ throughout the entire mass range.}}
\end{figure}

\section{Summary}

The discovery of BHs via the GWs emitted when they merge in a binary
has confirmed one of the milestone predictions of the Theory of
General Relativity, while at the same time opening a new window into
our exploration of the Universe. As often happens with new
observations, this new window has also raised some questions. In the
case of the 20 BHs discovered by the LIGO/Virgo collaboration via
their mergers in binaries during the O1/O2 runs, their mass spectrum has been somewhat  surprising,
being shifted towards larger masses with respect to the
BHs whose masses had previously been measured dynamically in X-ray
binaries.
{  The measured spins, on the other hand, have been found to be mostly consistent
with low and isotropic, unlike the generally high ones measured in X-ray binaries.
}

{  In this {\em Letter} we have investigated the possibility that  }
X-ray and GW-detected BHs are dominated by
different formation channels: isolated binary evolution for the former
and (primarily) dynamical formation for the latter.
Via high-resolution N-body simulations of a mini-cluster of initially
isolated BHs { (which can be thought of as the remnants of an OB
  association)}, we have shown a tendency for binary BH
formation among the heaviest objects in the cluster (see also
\citealt{OLeary2006,Rodriguez2018,Dicarlo2019} for similar trends in
globular clusters). The heaviest BH binaries tend to also be the ones
which are more tightly bound, hence resulting in shorter merger times,
which enhances the probability of being detected via GWs.

While weighed towards larger masses, the precise shape of the mass
distribution of the dynamically-formed binary BHs is also reflective
of the particular initial BH mass function.  {  We investigated this
 distribution for a variety of BH mass spectra, from a flat distribution to a powerlaw with index
 of -4 and extracted
 the sub-population of dynamically-formed binaries which merge within a Hubble time.
 We hence performed a Bayesian statistical analysis to compare the likelihood of each
 of these models to the LIGO/Virgo data from the O1 and O2 observing runs. We found that
 an initial BH mass spectrum $\propto M^{-2.35}$ is favored by the data, together with a maximum BH mass $M_{\rm max}\sim 50 M_\odot$. This is consistent with the theoretical upper limit for stellar BH masses, which is set by the occurrence of Pair Instability \citep{Woosley2017,Marchant2019}.

 A slope $\propto M^{-2.35}$  reflects the initial mass function of massive stars, which is expected
 at  the low metallicities required to form very massive BH remnants (i.e.  \citealt{Spera2015}).
 Hence our work shows that dynamically formed binaries from low-metallicity stars are
 reasonably compatible with the binary BH mass distribution  from the O1 and O2  LIGO/Virgo observing runs.

 As more data is expected to be gathered in the years to come,  statistical
 comparisons with numerical simulations
 will allow to establish whether the dynamical
 formation channel is  indeed dominant, and } to reconstruct the mass
spectrum of the initial BHs, thus shedding a new light on massive
stars and their evolution.


\acknowledgments
We thank  Johan Samsing for valuable comments on our manuscript.}
RP acknowledges support from the NSF under grant AST-1616157.
The Center for Computational Astrophysics at the Flatiron Institute is supported by the Simons Foundation.

\bibliographystyle{aasjournal}
\bibliography{biblio.bib}

\begin{thebibliography}{}
\expandafter\ifx\csname natexlab\endcsname\relax\def\natexlab#1{#1}\fi
\providecommand{\url}[1]{\href{#1}{#1}}
\providecommand{\dodoi}[1]{doi:~\href{http://doi.org/#1}{\nolinkurl{#1}}}
\providecommand{\doeprint}[1]{\href{http://ascl.net/#1}{\nolinkurl{http://ascl.net/#1}}}
\providecommand{\doarXiv}[1]{\href{https://arxiv.org/abs/#1}{\nolinkurl{https://arxiv.org/abs/#1}}}

\bibitem[{{Abbott} {et~al.}(2016{\natexlab{a}}){Abbott}, {Abbott}, {Abbott},
  {Abernathy}, {Ackley}, {Adams}, {Adams}, {Addesso}, {Adhikari}, \&
  et~al.}]{Abbott2016}
{Abbott}, B.~P., {Abbott}, R., {Abbott}, T.~D., {et~al.} 2016{\natexlab{a}},
  Physical Review Letters, 116, 061102, \dodoi{10.1103/PhysRevLett.116.061102}

\bibitem[{{Abbott} {et~al.}(2016{\natexlab{b}}){Abbott}, {Abbott}, {Abbott},
  {Abernathy}, {Acernese}, {Ackley}, {Adams}, {Adams}, {Addesso}, {Adhikari},
  {Adya}, {Affeldt}, {Agathos}, {Agatsuma}, {Aggarwal}, {Aguiar}, {Aiello},
  {Ain}, {Ajith}, {Allen}, {Allocca}, {Altin}, {Anderson}, {Anderson}, {Arai},
  {Araya}, {et~al.}}]{GW150914Astrophysics}
---. 2016{\natexlab{b}}, \apj, 818, L22, \dodoi{10.3847/2041-8205/818/2/L22}

\bibitem[{{Abbott} {et~al.}(2016{\natexlab{c}}){Abbott}, {Abbott}, {Abbott},
  {Abernathy}, {Acernese}, {Ackley}, {Adams}, {Adams}, {Addesso}, {Adhikari},
  {Adya}, {Affeldt}, {Agathos}, {Agatsuma}, {Aggarwal}, {Aguiar}, {Aiello},
  {Ain}, {Ajith}, {Allen}, {Allocca}, {Altin}, {Anderson}, {Anderson}, {Arai},
  {Araya}, {Arceneaux}, {Areeda}, {Arnaud}, {et~al.}}]{Abbott2016c}
---. 2016{\natexlab{c}}, The Astrophysical Journal Supplement Series, 227, 14,
  \dodoi{10.3847/0067-0049/227/2/14}

\bibitem[{{Antonini} {et~al.}(2016){Antonini}, {Chatterjee}, {Rodriguez},
  {Morscher}, {Pattabiraman}, {Kalogera}, \& {Rasio}}]{Antonini2016}
{Antonini}, F., {Chatterjee}, S., {Rodriguez}, C.~L., {et~al.} 2016, \apj, 816,
  65, \dodoi{10.3847/0004-637X/816/2/65}

\bibitem[{{Antonini} {et~al.}(2018){Antonini}, {Gieles}, \&
  {Gualandris}}]{Antonini2018}
{Antonini}, F., {Gieles}, M., \& {Gualandris}, A. 2018, arXiv e-prints.
\newblock \doarXiv{1811.03640}

\bibitem[{{Banerjee}(2018)}]{Banerjee2018}
{Banerjee}, S. 2018, \mnras, 473, 909, \dodoi{10.1093/mnras/stx2347}

\bibitem[{{Belczynski} {et~al.}(2016){Belczynski}, {Holz}, {Bulik}, \&
  {O'Shaughnessy}}]{Belczynski2016}
{Belczynski}, K., {Holz}, D.~E., {Bulik}, T., \& {O'Shaughnessy}, R. 2016,
  \nat, 534, 512, \dodoi{10.1038/nature18322}

\bibitem[{{Chatterjee} {et~al.}(2017){Chatterjee}, {Rodriguez}, \&
  {Rasio}}]{Chatterjee2017}
{Chatterjee}, S., {Rodriguez}, C.~L., \& {Rasio}, F.~A. 2017, \apj, 834, 68,
  \dodoi{10.3847/1538-4357/834/1/68}

\bibitem[{{Demorest} {et~al.}(2010){Demorest}, {Pennucci}, {Ransom}, {Roberts},
  \& {Hessels}}]{Demorest2010}
{Demorest}, P.~B., {Pennucci}, T., {Ransom}, S.~M., {Roberts}, M.~S.~E., \&
  {Hessels}, J.~W.~T. 2010, \nat, 467, 1081, \dodoi{10.1038/nature09466}

\bibitem[{{Di Carlo} {et~al.}(2019){Di Carlo}, {Giacobbo}, {Mapelli},
  {Pasquato}, {Spera}, {Wang}, \& {Haardt}}]{Dicarlo2019}
{Di Carlo}, U.~N., {Giacobbo}, N., {Mapelli}, M., {et~al.} 2019, arXiv
  e-prints.
\newblock \doarXiv{1901.00863}

\bibitem[{{Dominik} {et~al.}(2013){Dominik}, {Belczynski}, {Fryer}, {Holz},
  {Berti}, {Bulik}, {Mandel}, \& {O'Shaughnessy}}]{Dominik2013}
{Dominik}, M., {Belczynski}, K., {Fryer}, C., {et~al.} 2013, \apj, 779, 72,
  \dodoi{10.1088/0004-637X/779/1/72}

\bibitem[{{Farr}(2019)}]{Farr2019}
{Farr}, W.~M. 2019, arXiv e-prints, arXiv:1904.10879.
\newblock \doarXiv{1904.10879}

\bibitem[{{Farr} {et~al.}(2017){Farr}, {Stevenson}, {Miller}, {Mandel}, {Farr},
  \& {Vecchio}}]{Farr2017}
{Farr}, W.~M., {Stevenson}, S., {Miller}, M.~C., {et~al.} 2017, \nat, 548, 426,
  \dodoi{10.1038/nature23453}

\bibitem[{{Fishbach} {et~al.}(2018){Fishbach}, {Holz}, \&
  {Farr}}]{Fishback2018}
{Fishbach}, M., {Holz}, D.~E., \& {Farr}, W.~M. 2018, \apjl, 863, L41,
  \dodoi{10.3847/2041-8213/aad800}

\bibitem[{{Fragione} {et~al.}(2018){Fragione}, {Grishin}, {Leigh}, {Perets}, \&
  {Perna}}]{Fragione2018b}
{Fragione}, G., {Grishin}, E., {Leigh}, N.~W.~C., {Perets}, H.~B., \& {Perna},
  R. 2018, arXiv e-prints.
\newblock \doarXiv{1811.10627}

\bibitem[{{Fragione} \& {Kocsis}(2018)}]{Fragione2018a}
{Fragione}, G., \& {Kocsis}, B. 2018, Physical Review Letters, 121, 161103,
  \dodoi{10.1103/PhysRevLett.121.161103}

\bibitem[{{Generozov} {et~al.}(2018){Generozov}, {Stone}, {Metzger}, \&
  {Ostriker}}]{Generozov2018}
{Generozov}, A., {Stone}, N.~C., {Metzger}, B.~D., \& {Ostriker}, J.~P. 2018,
  \mnras, 478, 4030, \dodoi{10.1093/mnras/sty1262}

\bibitem[{{Harris}(1996)}]{Harris1996}
{Harris}, W.~E. 1996, \aj, 112, 1487, \dodoi{10.1086/118116}

\bibitem[{{Kalogera}(2000)}]{Kalogera2000}
{Kalogera}, V. 2000, \apj, 541, 319, \dodoi{10.1086/309400}

\bibitem[{{Leigh} {et~al.}(2014){Leigh}, {L{\"u}tzgendorf}, {Geller},
  {Maccarone}, {Heinke}, \& {Sesana}}]{Leigh2014}
{Leigh}, N.~W.~C., {L{\"u}tzgendorf}, N., {Geller}, A.~M., {et~al.} 2014,
  \mnras, 444, 29, \dodoi{10.1093/mnras/stu1437}

\bibitem[{{Madau} \& {Dickinson}(2014)}]{Madau2014}
{Madau}, P., \& {Dickinson}, M. 2014, Annual Review of Astronomy and
  Astrophysics, 52, 415, \dodoi{10.1146/annurev-astro-081811-125615}

\bibitem[{{Mandel} \& {de Mink}(2016)}]{Mandel2016}
{Mandel}, I., \& {de Mink}, S.~E. 2016, \mnras, 458, 2634,
  \dodoi{10.1093/mnras/stw379}

\bibitem[{{Mandel} {et~al.}(2018){Mandel}, {Farr}, \& {Gair}}]{Mandel2018}
{Mandel}, I., {Farr}, W.~M., \& {Gair}, J.~R. 2018, arXiv e-prints.
\newblock \doarXiv{1809.02063}

\bibitem[{{Mapelli} {et~al.}(2013){Mapelli}, {Zampieri}, {Ripamonti}, \&
  {Bressan}}]{Mapelli2013}
{Mapelli}, M., {Zampieri}, L., {Ripamonti}, E., \& {Bressan}, A. 2013, \mnras,
  429, 2298, \dodoi{10.1093/mnras/sts500}

\bibitem[{{Marchant} {et~al.}(2016){Marchant}, {Langer}, {Podsiadlowski},
  {Tauris}, \& {Moriya}}]{Merchant2016}
{Marchant}, P., {Langer}, N., {Podsiadlowski}, P., {Tauris}, T.~M., \&
  {Moriya}, T.~J. 2016, \aap, 588, A50, \dodoi{10.1051/0004-6361/201628133}

\bibitem[{{Marchant} {et~al.}(2018){Marchant}, {Renzo}, {Farmer}, {Pappas},
  {Taam}, {de Mink}, \& {Kalogera}}]{Marchant2019}
{Marchant}, P., {Renzo}, M., {Farmer}, R., {et~al.} 2018, arXiv e-prints.
\newblock \doarXiv{1810.13412}

\bibitem[{{Mikkola} \& {Merritt}(2008)}]{Mikkola2008}
{Mikkola}, S., \& {Merritt}, D. 2008, \aj, 135, 2398,
  \dodoi{10.1088/0004-6256/135/6/2398}

\bibitem[{{Miller} \& {Lauburg}(2009)}]{Miller2009}
{Miller}, M.~C., \& {Lauburg}, V.~M. 2009, \apj, 692, 917,
  \dodoi{10.1088/0004-637X/692/1/917}

\bibitem[{{O'Leary} {et~al.}(2006){O'Leary}, {Rasio}, {Fregeau}, {Ivanova}, \&
  {O'Shaughnessy}}]{OLeary2006}
{O'Leary}, R.~M., {Rasio}, F.~A., {Fregeau}, J.~M., {Ivanova}, N., \&
  {O'Shaughnessy}, R. 2006, \apj, 637, 937, \dodoi{10.1086/498446}

\bibitem[{{Peters}(1964)}]{Peters1964}
{Peters}, P.~C. 1964, Physical Review, 136, 1224,
  \dodoi{10.1103/PhysRev.136.B1224}

\bibitem[{{Podsiadlowski} {et~al.}(2003){Podsiadlowski}, {Rappaport}, \&
  {Han}}]{Podsiadlowski2003}
{Podsiadlowski}, P., {Rappaport}, S., \& {Han}, Z. 2003, \mnras, 341, 385,
  \dodoi{10.1046/j.1365-8711.2003.06464.x}

\bibitem[{{Portegies Zwart} \& {McMillan}(2000)}]{Portegies2000}
{Portegies Zwart}, S.~F., \& {McMillan}, S.~L.~W. 2000, \apjl, 528, L17,
  \dodoi{10.1086/312422}

\bibitem[{{Rodriguez} {et~al.}(2018){Rodriguez}, {Amaro-Seoane}, {Chatterjee},
  \& {Rasio}}]{Rodriguez2018a}
{Rodriguez}, C.~L., {Amaro-Seoane}, P., {Chatterjee}, S., \& {Rasio}, F.~A.
  2018, Physical Review Letters, 120, 151101,
  \dodoi{10.1103/PhysRevLett.120.151101}

\bibitem[{{Rodriguez} {et~al.}(2016){Rodriguez}, {Haster}, {Chatterjee},
  {Kalogera}, \& {Rasio}}]{Rodriguez2016}
{Rodriguez}, C.~L., {Haster}, C.-J., {Chatterjee}, S., {Kalogera}, V., \&
  {Rasio}, F.~A. 2016, \apjl, 824, L8, \dodoi{10.3847/2041-8205/824/1/L8}

\bibitem[{{Rodriguez} \& {Loeb}(2018)}]{Rodriguez2018}
{Rodriguez}, C.~L., \& {Loeb}, A. 2018, \apjl, 866, L5,
  \dodoi{10.3847/2041-8213/aae377}

\bibitem[{{Roulet} \& {Zaldarriaga}(2019)}]{Roulet2019}
{Roulet}, J., \& {Zaldarriaga}, M. 2019, \mnras, \dodoi{10.1093/mnras/stz226}

\bibitem[{{Ryu} {et~al.}(2016){Ryu}, {Tanaka}, \& {Perna}}]{Ryu2016}
{Ryu}, T., {Tanaka}, T.~L., \& {Perna}, R. 2016, \mnras, 456, 223,
  \dodoi{10.1093/mnras/stv2629}

\bibitem[{{Salpeter}(1955)}]{Salpeter1955}
{Salpeter}, E.~E. 1955, \apj, 121, 161, \dodoi{10.1086/145971}

\bibitem[{{Samsing}(2018)}]{Samsing2018}
{Samsing}, J. 2018, \prd, 97, 103014, \dodoi{10.1103/PhysRevD.97.103014}

\bibitem[{{Samsing} \& {D'Orazio}(2018)}]{Samsing2018a}
{Samsing}, J., \& {D'Orazio}, D.~J. 2018, \mnras, 481, 5445,
  \dodoi{10.1093/mnras/sty2334}

\bibitem[{{Sigurdsson} \& {Hernquist}(1993)}]{Sigurdsson1993}
{Sigurdsson}, S., \& {Hernquist}, L. 1993, \nat, 364, 423,
  \dodoi{10.1038/364423a0}

\bibitem[{{Spera} {et~al.}(2015){Spera}, {Mapelli}, \& {Bressan}}]{Spera2015}
{Spera}, M., {Mapelli}, M., \& {Bressan}, A. 2015, \mnras, 451, 4086,
  \dodoi{10.1093/mnras/stv1161}

\bibitem[{{The LIGO Scientific Collaboration} \& {the Virgo
  Collaboration}(2018)}]{Abbott2019}
{The LIGO Scientific Collaboration}, \& {the Virgo Collaboration}. 2018, ArXiv
  e-prints, arXiv:1811.12907.
\newblock \doarXiv{1811.12907}

\bibitem[{{The LIGO Scientific Collaboration} {et~al.}(2018){The LIGO
  Scientific Collaboration}, {the Virgo Collaboration}, {Abbott}, {Abbott},
  {Abbott}, {Abraham}, {Acernese}, {Ackley}, {Adams}, {Adhikari}, \&
  et~al.}]{Abbott2018}
{The LIGO Scientific Collaboration}, {the Virgo Collaboration}, {Abbott},
  B.~P., {et~al.} 2018, arXiv e-prints.
\newblock \doarXiv{1811.12940}

\bibitem[{{Wang} {et~al.}(2019){Wang}, {Leigh}, {Sesana}, \&
  {Perna}}]{Wang2019}
{Wang}, Y.-H., {Leigh}, N., {Sesana}, A., \& {Perna}, R. 2019, \mnras, 482,
  3206, \dodoi{10.1093/mnras/sty2866}

\bibitem[{{Wiktorowicz} {et~al.}(2014){Wiktorowicz}, {Belczynski}, \&
  {Maccarone}}]{Witz2014}
{Wiktorowicz}, G., {Belczynski}, K., \& {Maccarone}, T. 2014, in Binary
  Systems, their Evolution and Environments, 37

\bibitem[{{Woosley}(2017)}]{Woosley2017}
{Woosley}, S.~E. 2017, \apj, 836, 244, \dodoi{10.3847/1538-4357/836/2/244}

\bibitem[{{Ye} {et~al.}(2019){Ye}, {Kremer}, {Chatterjee}, {Rodriguez}, \&
  {Rasio}}]{Ye2019}
{Ye}, C.~S., {Kremer}, K., {Chatterjee}, S., {Rodriguez}, C.~L., \& {Rasio},
  F.~A. 2019, arXiv e-prints.
\newblock \doarXiv{1902.05963}

\bibitem[{{Zhou} {et~al.}(1994){Zhou}, {Evans}, {Wang}, {Peng}, \&
  {Lo}}]{Zhou1994}
{Zhou}, S., {Evans}, II, N.~J., {Wang}, Y., {Peng}, R., \& {Lo}, K.~Y. 1994,
  \apj, 433, 131, \dodoi{10.1086/174630}

\bibitem[{{Ziosi} {et~al.}(2014){Ziosi}, {Mapelli}, {Branchesi}, \&
  {Tormen}}]{Ziosi2014}
{Ziosi}, B.~M., {Mapelli}, M., {Branchesi}, M., \& {Tormen}, G. 2014, \mnras,
  441, 3703, \dodoi{10.1093/mnras/stu824}

\end{thebibliography}

\end{document}